\documentstyle[aps,prb,twocolumn]{revtex}
\draft
\begin{document}
\def\vec#1{{\bf{#1}}}
\title{Transport theory yields renormalization group equations}
\author{Jochen Rau}
\address{Max-Planck-Institut f\"ur Physik komplexer Systeme,
Bayreuther Stra{\ss}e 40 Haus 16,
01187 Dresden, Germany}
\date{revised version, January 21, 1997; to appear in Phys. Rev. E}
\maketitle
\begin{abstract}
We show that dissipative transport and renormalization can be
described in a single theoretical framework.
The appropriate mathematical tool is the Nakajima-Zwanzig
projection technique.
We illustrate our result in the case of interacting quantum gases,
where we use the Nakajima-Zwanzig approach to investigate the
renormalization group flow of the effective two-body interaction.
\end{abstract}
\medskip
\narrowtext
\section{Introduction}
The basic theme of statistical mechanics --how to obtain a system's
macroscopic properties from the laws of its underlying microscopic
dynamics-- appears in many variations.
Two out of many examples are
the problem of determining critical exponents at second-order
phase transitions,
or the problem of deriving macroscopic transport equations.
The former is usually tackled with the help of
Wilson's renormalization group,\cite{kadanoff,wilson,wegner,polchinski}
a mathematical tool that allows one to 
iteratively eliminate short-wavelength modes and thus to arrive at
effective (``renormalized'')
theories which describe the dynamics on successively
larger length scales.
The latter has been tackled in various ways, among them
the so-called projection technique by 
Nakajima,\cite{nakajima} Zwanzig,\cite{zwanzig} 
Mori\cite{mori} and 
Robertson.\cite{robertson} 
Eliminating
unmonitored, rapidly oscillating degrees of freedom
from the equation of motion by means of suitable projections
in the space of observables,
the projection technique yields closed (but generally no longer
Markovian) ``transport equations'' for the selected macroscopic
degrees of freedom.

While the two methods
--Wilson's renormalization group and Zwanzig's projection technique--
may appear quite different in their mathematical formulation,
they are very similar in spirit.
In both cases one strives to focus on selected features of the
dynamics (its infrared limit, or the evolution of only few
macroscopic observables) deemed interesting, and to this end
devises a systematic procedure for
eliminating all other, ``irrelevant'' degrees of freedom
(a procedure commonly referred to as ``coarse graining'').
Discarding thus unnecessary ``baggage'' from the
problem at hand, one succeeds in describing the interesting
features of the dynamics without ever having to solve
the complete, and far too complicated,
microscopic theory.
This similarity of the basic approach suggests that renormalization and the
transition from microdynamics to macroscopic transport are in fact
closely related procedures, and that it should be possible to
cast them into a common theoretical framework.

Building a bridge between renormalization and
transport theory would be not only satisfying conceptually,
but would also help tackle a
variety of practical problems.
Often, the macroscopic evolution of a complex quantum system
exhibits {\em both} dissipation {\em and} modified, ``renormalized''
dynamical parameters such as effective masses or effective
interactions.
Let us consider, for example, liquid
$^3\mbox{He}$ or nuclear matter away from equilibrium.
In order to formulate a macroscopic transport theory for such an interacting
fermion system one must perform
{\em two} consecutive coarse graining procedures:
first eliminating short-wavelength modes to arrive at an
effective (``renormalized'')
theory for quasiparticle excitations close to the Fermi surface,
which typically feature effective masses and 
screened interactions;\cite{polchinski_tasi,weinberg,shankar}
and then discarding their statistical correlations to obtain
an Uehling-Uhlenbeck-type transport equation for the single-quasiparticle
distribution.
The description of the macroscopic dynamics, therefore,
requires an appropriate combination of renormalization and 
statistical coarse graining.

Clearly, the two coarse grainings do not commute;
the latter (statistical coarse graining) is contingent upon
the former (renormalization).
For instance, the renormalization group flow yields screening\cite{shankar}
and hence renders the interaction range finite,
thus generating that separation of scales 
which is indispensable 
for the subsequent derivation of a
Markovian transport theory.\cite{rau}
But what happens if scales converge rather than separate
upon renormalization? 
How are then renormalization and statistical coarse graining
best combined?
More generally, what is the connection between 
effective dynamics and dissipation?
Does their interplay lead to interesting new phenomena?
To what extent can renormalization group techniques be applied to study
nonequilibrium, dissipative processes?
How do transport coefficients change under renormalization
group transformations?
Is it always true that transport coefficients are renormalized by
simply trading bare masses and couplings for their
renormalized counterparts, while keeping the form of the functional
dependence on these parameters?\cite{jeon:yaffe}
And, somewhat speculative, are there ``universality classes'' 
of transport theories?
These and other issues might be best approached in a 
unified mathematical framework 
that encompasses both
renormalization and dissipative transport as special cases.

There has already been some progress towards such a unified picture.
The success of Anderson's 
``poor man's scaling'' approach to the Kondo problem,\cite{anderson}
Seke's projection-method treatment of the nonrelativistic
Lamb shift,\cite{seke}
the calculation of the 1-loop renormalization of $\phi^4$ theory
by means of Bloch-Feshbach techniques,\cite{muller:rau}
or a recent renormalization group study of interacting fermion systems 
(BCS instability, screening)
within a purely algebraic framework,\cite{rau_prb}
suggest that one can formulate Wilson's renormalization 
in terms of projections in Hilbert space,
completely analogous to the projections in the space
of observables which, in the Nakajima-Zwanzig approach, would lead to
macroscopic transport equations.

In the present paper I wish to make this analogy
even more explicit.
I will show that one can actually obtain
renormalization group equations within the
Nakajima-Zwanzig projection approach,
and that hence renormalization can be embedded into the
general mathematical framework of transport theory.
After a brief introduction to the projection technique
(Sec. \ref{technique})
and a discussion of various approximations
(Sec. \ref{approximations})
I shall
isolate the dissipative and non-dissipative parts of the 
macroscopic dynamics,
and show that the latter is
governed by an effective, ``renormalized'' Hamiltonian
(Sec. \ref{analysis}).
For illustration, these general ideas are then applied to studying
the low-energy dynamics of interacting quantum (Bose and Fermi) gases,
in particular the renormalization group flow
of their effective two-body interaction
(Secs. \ref{gases} and \ref{examples}).
Finally, I shall conclude with a brief summary
in Sec. \ref{summary}.
\section{Projection technique}\label{technique}
In this section I give a very brief introduction to the
Nakajima-Zwanzig projection
technique.\cite{nakajima,zwanzig,mori,robertson} 
More details can be found in several 
textbooks\cite{textbooks} and in recent reviews.\cite{rau,balian}

When studying the dynamics of a macroscopic quantum system
away from equilibrium,
one typically monitors the evolution of the 
expectation values
\begin{equation}\label{exvalues}
g_a(t):=
\mbox{tr}[\rho(t)G_a]
\end{equation}
of only a very small set of selected
(``relevant'')
observables $\{G_a\}$.
These evolve according to
\begin{equation}\label{lvn}
\dot g_a(t)=i(\rho(t)|{\cal L}G_a)
\quad,
\end{equation}
with $\rho(t)$ being the statistical operator,
${\cal L}$ the Liouvillian
\begin{equation}
{\cal L}:=\hbar^{-1}[H,*]
\end{equation}
associated with the Hamiltonian $H$,
and the inner product $(\cdot|\cdot)$ defined as
\begin{equation}
(A|B):=\mbox{tr}[A^\dagger B]
\quad.
\end{equation}
The equation of motion in the form (\ref{lvn}) 
does not constitute a closed system
of differential equations for the selected
expectation values $\{g_a(t)\}$;
its right-hand side will generally 
depend not just on the selected, but also on all the other
(``irrelevant'') degrees of freedom.
With the help of the projection technique to be sketched below,
these irrelevant degrees of freedom can be systematically
eliminated from the equation
of motion, in exchange for non-Markovian and (possibly) nonlinear
features of the resulting closed
``transport equation'' for the $\{g_a(t)\}$.
Mapping thus the influence of irrelevant degrees of freedom 
onto --among other features-- a non-local
behavior in time opens the way to the exploitation of 
well-separated time scales,
and hence serves as a good
starting point for powerful approximations such as
the Markovian and quasistationary limits.
Indeed, in this fashion one can derive 
many of the well-known equations of nonequilibrium
statistical mechanics,
for example
rate equations, the quantum Boltzmann, Master,
Langevin-Mori or even
time-dependent Hartree-Fock equations.
And as I will show later in this paper, the same projection 
technique allows one to derive
renormalization group equations.

The projection technique
is based on a clever insertion of projection operators
into the equation of motion (\ref{lvn}).
A projection operator is any operator ${\cal P}$ which
satisfies ${\cal P}^2={\cal P}$; its complement,
which is also a projection operator, is denoted by
${\cal Q}:=1-{\cal P}$.
The projection operators, like the Liouvillian, are so-called
``superoperators:'' they do not act in 
Hilbert space, but in the space of observables 
(``Liouville space'').
For our purposes we consider projectors which
project arbitrary vectors in Liouville space onto the 
subspace spanned by the unit operator and by the 
relevant observables $\{G_a\}$;
i.e., for which
\begin{equation}
{\cal P} A= A
\quad\Leftrightarrow\quad 
A\in {\rm span}\{1,{G_a}\}
\quad.
\label{project}
\end{equation}
For simplicity we assume that the Hamiltonian and hence the
Liouvillian, as well as 
the relevant observables,
are not explicitly time-dependent.
In contrast, we allow the projector to depend 
on the expectation values $\{g_a(t)\}$ of the relevant observables, 
thus making it
an implicit function of time: ${\cal P}(t)\equiv {\cal P}[g_a(t)]$;
with the sole restriction that for any observable $A$,
\begin{equation}
({\rho(t)}|{\mbox{d}\over\mbox{d} t}{\cal P}(t)\,A)=0
\quad.
\label{technical}
\end{equation}
For the time being we admit any projector which satisfies the two
constraints (\ref{project}) and (\ref{technical}).
Only later, in Sec. \ref{sec_robertson}, we shall make a specific
choice for ${\cal P}(t)$.

Now let
${\cal T}(t',t)$ be the (super)operator defined by
the differential equation
\begin{equation}
\label{irrel:evolution}
{\partial\over\partial t'}{\cal T}(t',t)=
-{i}\,{\cal Q}(t'){\cal L}{\cal Q}(t')
{\cal T}(t',t) 
\end{equation}
with initial condition
${\cal T}(t,t)=1$.
It may be pictured as describing
the evolution of the system's {\em irrelevant} degrees of freedom.
With its help
the equation of motion
for the selected expectation values $\{g_a(t)\}$
can be cast into the form
\begin{eqnarray}
\dot g_a (t)
&=& 
{i}(\rho(t)|{\cal P}(t){\cal L}{G_a})
\nonumber \\
&-&
\int_{0}^t\mbox{d} t'\,
(\rho(t')|{\cal P}(t'){\cal L}{\cal Q}(t')
{\cal T}(t',t){\cal Q}(t){\cal L}{G_a})
\nonumber \\
&+& 
{i}(\rho(0)|{\cal Q}(0)
{\cal T}(0,t){\cal Q}(t){\cal L}{G_a})
\quad,
\label{robertson_2}
\end{eqnarray}
for any time $t\ge 0$.
Comparing this form of the equation of motion with
the original form (\ref{lvn})
we notice that, apart from the replacement $(\rho|\to (\rho|{\cal P}$,
there are two terms which are qualitatively new:
(i) an integral (``memory'') term,
containing contributions from all times between the initial and
the present time; 
and (ii) a ``residual force'' term describing the effect
of irrelevant components in the initial state.
The physical meaning of both terms can be easily
discerned if read from left to right:
(i) At time $t'<t$ relevant degrees of freedom
(projected out by ${\cal P}$) couple via an interaction (${\cal L}$)
to irrelevant degrees of freedom (projected out by ${\cal Q}$),
which subsequently evolve in time (${\cal T}$) and,
due to a second interaction (${\cal L}$), acquire relevancy again, 
thus influencing the evolution of the relevant observable
$G_a$ at
the present time $t$.
(ii) Irrelevant components in the initial state (${\cal Q}$) evolve in time
(${\cal T}$) and, due to interaction (${\cal L}$),
acquire relevancy at the present time $t$.

In many practical applications the irrelevant component of
the initial state can be shown to vanish, or else to be
negligible; in which case the residual force term can be
dropped from the equation of motion.
What remains, is then the desired closed 
system of (possibly nonlinear) coupled integro-differential equations
for the selected expectation values 
$\{g_a(t)\}$.
The principal feature of these closed equations 
is that they are non-Markovian:
future expectation values of the selected observables are
predicted not just on the basis of their present values,
but based on their entire history.
\section{Approximations}\label{approximations}
\subsection{Second order perturbation theory}
Often the Liouvillian can be split into a free part and
an interaction part,
\begin{equation}
{\cal L}={\cal L}^{(0)}+{\cal V}
\quad,
\end{equation}
corresponding to a decomposition $H=H^{(0)}+ V$
of the Hamiltonian.
Provided free evolution does not mix relevant and irrelevant
degrees of freedom, i.e., provided
\begin{equation}
[{{\cal L}^{(0)}},{{\cal P}(t)}]=0
\quad,
\label{commute}
\end{equation}
then in the memory term ${\cal PLQ}={\cal PVQ}$ and
${\cal QLP}={\cal QVP}$; hence
the memory term is at least of second order in
the interaction.
To second order, therefore, one can simply replace
\begin{equation}
{\cal Q}(t'){\cal T}(t',t){\cal Q}(t)
\,\to\,
{\cal Q}(t'){\cal T}^{(0)}(t',t){\cal Q}(t)
\quad,
\end{equation}
where ${\cal T}^{(0)}$ is a time-ordered exponential
of ${\cal Q}{\cal L}^{(0)}{\cal Q}$.
Using (\ref{commute}) and ${\cal Q}(t_2){\cal Q}(t_1)={\cal Q}(t_2)$,
all the ${\cal Q}$'s appearing in ${\cal T}^{(0)}$
can be shuffled to the left
and absorbed into ${\cal Q}(t')$, allowing one to replace further
\begin{equation}
{\cal T}^{(0)}(t',t)
\,\to\,
{\cal U}^{(0)}(t',t):=\exp[i{\cal L}^{(0)}\cdot (t-t')]
\quad.
\end{equation}
One thus obtains (without residual force)
\begin{eqnarray}
\dot{g_a}(t)
&=&
{i}({\rho(t)}|{\cal P}(t){\cal L}{G_a})
\nonumber \\
&-&
\int_0^{t}\!\mbox{d}\tau \, ({\rho(t-\tau)}|
{\cal P}(t-\tau){\cal V}
{\cal Q}(t-\tau){\cal U}^{(0)}(0,\tau)
{\cal V}{G_a})
\, .
\nonumber \\
&&\mbox{}
\label{secorder}
\end{eqnarray}
\subsection{Markovian limit}
We have seen that predictions of future expectation
values of the selected observables generally depend
in a complicated manner on both their present expectation values
and their past history.
There are thus two distinct time scales:
(i) the scale $\tau_{\rm rel}$ --or several scales 
$\{\tau_{\rm rel}^{i}\}$-- on which the 
selected expectation values $\{g_a(t)\}$
evolve;
and (ii) the {\em memory time} $\tau_{\rm mem}$, which
characterizes the length of the time interval that contributes
significantly to the memory integral. 
Loosely speaking, the memory time determines how
far back into the past one has to reach
in order to make predictions 
for the further evolution of the selected observables.
If this memory time is small compared to
the typical time scale on which the selected observables evolve,
$\tau_{\rm mem}\ll\tau_{\rm rel}$, 
then memory effects can be neglected and predictions for the
selected observables be based solely on their present values.
One may then assume that
in the memory term $g_a(t')\approx g_a(t)$ and hence
replace
\begin{eqnarray}
{\cal P}[g_a(t')]
&\to&
{\cal P}[g_a(t)]
\quad,
\nonumber \\
(\rho(t')|{\cal P}(t')
&\to&
(\rho(t)|{\cal P}(t)
\quad.
\end{eqnarray}
This is the Markovian limit.

Closely related to the Markovian limit is the
quasistationary limit:
at times $t\gg\tau_{\rm mem}$ it no longer matters 
for the dynamics when exactly
the evolution started, and hence in Eq. (\ref{robertson_2})
the integration over the
system's history may just as well extend from $-\infty$ to $t$,
rather than from $0$ to $t$.

In the Markovian and quasistationary limits
the equation of motion
--to second order and without residual force-- 
simplifies to
\begin{eqnarray}
\dot{g_a}(t)
&=&
{i}(\rho(t)|{\cal P}(t){\cal L}G_a)
\nonumber \\
&-&
\int_0^{\infty}\mbox{d}\tau\,
(\rho(t)|{\cal P}(t){\cal V}{\cal Q}(t)
{\cal U}^{(0)}(0,\tau)
{\cal V}G_a)
\, .
\label{approx}
\end{eqnarray}
This approximate transport equation for the $\{g_a(t)\}$ shall
be the basis of our further investigations.
\section{Analysis}\label{analysis}
\subsection{Reformulation of the transport equation}
Starting from the approximate transport equation (\ref{approx}),
we eventually wish to discern 
two of the main features of macroscopic transport:
(i) dissipation and (ii) the modification (``renormalization'')
of the effective interaction.
The latter will then, in Section \ref{gases}, lead on
to the consideration of
renormalization group equations.

To this end we first
split the free evolution operator into its symmetric
and antisymmetric parts,
\begin{eqnarray}
{\cal U}^{(0)}(0,\tau)
&=&
{\textstyle{1\over2}}
[{{\cal U}^{(0)}(0,\tau)+{\cal U}^{(0)}(\tau,0)}]
\nonumber \\
&&+
{\textstyle{1\over2}}
[{{\cal U}^{(0)}(0,\tau)-{\cal U}^{(0)}(\tau,0)}]
\quad,
\end{eqnarray}
use
\begin{equation}
\int_0^{\infty}\mbox{d}\tau\,
[{{\cal U}^{(0)}(0,\tau)+{\cal U}^{(0)}(\tau,0)}]
=2\pi\delta({\cal L}^{(0)})
\end{equation}
in the symmetric part,
and in the antisymmetric part
exploit the liberty --thanks to the Markovian limit--
to insert free evolution operators 
${\cal U}^{(0)}(0,\tau)$ and ${\cal U}^{(0)}(\tau,0)$ in
front of ${\cal P}(t)$ or $G_a$, respectively.
We thus obtain
\begin{eqnarray}
\dot{g_a}(t)
&=&
{i}(\rho(t)|{\cal P}(t){\cal L}_{\rm eff}(t)G_a)
\nonumber \\
&&-\pi
(\rho(t)|{\cal P}(t){\cal V}{\cal Q}(t)
\delta({\cal L}^{(0)})
{\cal V}G_a)
\quad,
\label{reformulated}
\end{eqnarray}
where
\begin{equation}
{\cal L}_{\rm eff}(t)={\cal L}+ \delta{\cal L}(t)
\end{equation}
denotes a (possibly time-dependent) ``effective'' Liouvillian
determined by
\begin{eqnarray}
\delta{\cal L}(t)=
-{i\over2}\int_0^\infty\! d\tau
&\{& 
[{\cal V}(\tau),{\cal V}]
\nonumber \\
&& -
[{\cal P}(t){\cal V}(\tau){\cal P}(t),{\cal P}(t){\cal V}{\cal P}(t)]
\}
\, ,
\label{delta_L}
\end{eqnarray}
and
\begin{equation}
{\cal V}(\tau)=
\hbar^{-1} [V(\tau), *]
\end{equation}
is the commutator
with the interaction-picture operator
\begin{equation}
V(\tau):={\cal U}^{(0)}(0,\tau) V
\quad.
\end{equation}
 
Next, we evaluate
the reformulated transport equation (\ref{reformulated}) 
by making a suitable choice for the
yet undetermined projector ${\cal P}(t)$.
\subsection{Robertson projector}\label{sec_robertson}
Associated with the 
expectation values (\ref{exvalues}) of the selected observables
is, at each time $t$,
a generalized canonical state
\begin{equation}\label{gen_canon}
\rho_{\rm rel}(t):=Z(t)^{-1}\exp[
- \lambda^a(t) G_a]
\quad,
\end{equation}
with partition function
\begin{equation}
Z(t):=\mbox{tr}\,\exp[
- \lambda^a(t) G_a]
\end{equation}
and the Lagrange parameters $\{\lambda^a(t)\}$ adjusted such
as to satisfy the constraints
\begin{equation}
\mbox{tr}[\rho_{\rm rel}(t) G_a]=g_a(t)
\quad.
\label{constraints}
\end{equation}
Here we have used Einstein's convention:
repeated upper and lower indices are to be summed over.
The generalized canonical state $\rho_{\rm rel}(t)$,
among all states which satisfy the constraints (\ref{constraints}),
is the one which maximizes
the von Neumann entropy
\begin{equation}
S[\rho]:=-k\,\mbox{tr}\,(\rho\ln\rho)
\quad.
\end{equation}
For this reason it may be
considered
``least biased'' or ``maximally non-committal'' 
with regard to the unmonitored degrees of freedom;
it is sometimes 
called the ``relevant part'' of the full statistical
operator $\rho(t)$.

There exists a unique time-dependent
projector ${\cal P}_{\rm R}(t)$ which projects
arbitrary vectors in Liouville space onto the
subspace spanned by the unit operator and by the 
relevant observables $\{G_a\}$, the projection being
orthogonal with respect to the time-dependent scalar product
\begin{equation}
\langle A;B\rangle^{(t)} :=
\int_0^1 d\mu\,\mbox{tr}\left[\rho_{\rm rel}(t)^\mu
A^\dagger \rho_{\rm rel}(t)^{1-\mu}B\right]
\quad.
\end{equation}
This projector ${\cal P}_{\rm R}(t)$  satisfies both
conditions (\ref{project}) and (\ref{technical}) and,
moreover, can be shown to yield
\begin{equation}
(\rho(t)|{\cal P}_{\rm R}(t)=(\rho_{\rm rel}(t)|
\end{equation}
at all times.
This special
choice for ${\cal P}(t)$, originally proposed by 
Robertson,\cite{robertson,newfoot}
has the important advantage that in contrast to other
frequently used projectors such as the Mori projector\cite{mori}
it permits the derivation of closed transport equations valid
arbitrarily far from equilibrium.
We shall use this projector throughout the remainder of the paper
(and for brevity, we shall immediately drop the subscript `R').

With the Robertson projector the transport equation
(\ref{reformulated}) takes the form
\begin{equation}
\dot{g_a}(t)
=
{i}(\rho_{\rm rel}(t)|{\cal L}_{\rm eff}(t)G_a)
+
M_{ca}(t) \lambda^c(t)
\quad,
\label{robertson}
\end{equation}
where
\begin{equation}
M_{ca}(t):=
\pi \langle{\cal Q}(t){\cal V}G_c;
\delta({\cal L}^{(0)}){\cal Q}(t){\cal V}G_a
\rangle^{(t)} 
\end{equation}
is a matrix whose eigenvalues are all
real and non-negative.
In this formulation it is particularly
easy to distinguish the dissipative and non-dissipative parts
of the dynamics:
as we will show,
the second term in
(\ref{robertson}) is solely responsible
for dissipation; whereas the first term yields non-dissipative
dynamics governed by a modified (``renormalized'') effective
Hamiltonian.
\subsection{Dissipation}
An observer who monitors only
the selected degrees of freedom does not have complete information
about the system's microstate.
A suitable measure for this lack of information 
is the entropy $S[\rho_{\rm rel}(t)]$
associated with the relevant part of the statistical operator.
It is sometimes called the ``relevant entropy.''
This relevant entropy generally varies in time:
it changes at a rate
\begin{equation}
\dot S[\rho_{\rm rel}(t)]=k\,\lambda^a(t)\dot g_a(t)
\quad.
\end{equation}
Within our approximations --perturbation theory
and Markovian limit-- 
this rate may be evaluated by inserting
the transport equation (\ref{robertson}).
Its first term does not contribute
to the change of relevant entropy; only its second term 
yields a nontrivial contribution
\begin{equation}
\dot S[\rho_{\rm rel}(t)]=
k\,M_{ca}(t)\lambda^c(t) \lambda^a(t)
\ge 0
\quad.
\end{equation}
The relevant entropy thus increases monotonically,
reflecting dissipation and irreversibility
of the macroscopic dynamics.
It stays constant if and only if the second
term in (\ref{robertson}) vanishes (``adiabatic limit'').

Of course, our finding represents one particular case of the
more general $H$-theorem. 
That the relevant entropy can never decrease,
is a direct consequence of the
Markovian limit and hence holds true whenever the system exhibits
a clear separation of time scales.\cite{rau}
\subsection{Effective Hamiltonian}
The non-dissipative part of the macroscopic dynamics
is encoded entirely in the first term of (\ref{robertson}).
So in the adiabatic limit, which we shall consider from now on,
the transport equation simplifies to
\begin{equation}
\dot{g_a}^{\rm ad}(t)
=
{i}(\rho_{\rm rel}(t)|{\cal L}_{\rm eff}(t)G_a)
\quad.
\end{equation}
This is
very similar to the original equation of motion
(\ref{lvn}), yet with $\rho$ replaced by $\rho_{\rm rel}$,
and ${\cal L}$ replaced by ${\cal L}_{\rm eff}$.

Here we make a special choice for the selected observables,
one that will directly lead to the renormalization group.
We presume that we are interested in
features of the macroscopic system (for example, its long-wavelength
properties) which can be represented
by observables acting merely in some subspace of the original
Hilbert space (e. g., in the subspace spanned by all many-particle states
with momenta below a given cutoff).
Let the operator which projects the 
original Hilbert space
onto this selected subspace be denoted by $P$, and 
its complement by $Q=1-P$.
Selected observables are then all those of the form
$PAP$, with $A$ being an arbitrary Hermitian operator.
This choice of relevant observables gives rise to a
particularly simple representation of the Robertson projector,
\begin{equation}
{\cal P}A = PAP + {\mbox{tr}(QA)\over\mbox{tr} Q} Q
\quad\forall\,\,A
\quad,
\end{equation}
which no longer varies in time.
 
We now decompose the microscopic Hamiltonian $H$
into a non-mixing ``free'' part
\begin{equation}
H^{(0)}=PHP+QHQ
\end{equation} 
and a mixing ``interaction''
\begin{equation}
V=PHQ+QHP
\quad,
\end{equation}
and split the Liouvillian correspondingly.
Since
\begin{equation}
[{\cal L}^{(0)},{\cal P}]=0
\quad,
\end{equation}
this decomposition is suitable for
perturbation theory.
With the property
\begin{equation}
{\cal PVP}=0
\end{equation}
we can immediately evaluate Eq. (\ref{delta_L}) to obtain
\begin{equation}
\delta{\cal L}=\hbar^{-1}[\delta H,*]
\quad,
\end{equation}
where $\delta H$ is given by
\begin{eqnarray}
\delta H
&=&
{i\over 2\hbar}\int_0^\infty d\tau\,
[V,V(\tau)]
\nonumber \\
&=&
-{1\over2\hbar} [V, {1\over{\cal L}^{(0)}} V]
+i\,{\pi\over2\hbar} [V, \delta({\cal L}^{(0)}) V]
\quad.
\end{eqnarray}
This yields then
\begin{equation}
{\cal P}{\cal L}_{\rm eff}{\cal P}A=
\hbar^{-1} [H_{\rm eff},PAP]
\quad\forall\,\,A
\quad,
\end{equation}
where we have identified the effective Hamiltonian
\begin{equation}
\label{h_eff}
H_{\rm eff}=PHP + \Sigma 
\end{equation}
which is not just the projection $PHP$ of the original
Hamiltonian, but contains an extra term
\begin{eqnarray}
\Sigma
&=&
- {1\over2\hbar} \left\{
PHQ {1\over{\cal L}^{(0)}} QHP + \mbox{h.c.}
\right\}
\nonumber \\
&&\quad
+ {\pi\over2\hbar} \{i PHQ \,\delta({\cal L}^{(0)})\, QHP +
\mbox{h.c.} \}
\label{sigma}
\end{eqnarray}
stemming from $\delta H$.
In $\Sigma$ the contributions which involve 
$\delta({\cal L}^{(0)})$ may be omitted
as long as the $P$- and $Q$-sectors of Hilbert space are 
associated with clearly distinct energies.

Our result for the effective Hamiltonian, 
which we have obtained within the
general framework of transport theory, 
is very similar to the
{\em Bloch-Feshbach formula} known in the theory
of nuclear dynamics,\cite{bloch}
or to Anderson's {\em poor man's scaling}.\cite{anderson}
Below we wish to demonstrate how this result
can be utilized to derive
renormalization group equations for a variety of
physical systems.
\section{Renormalization group for interacting quantum
gases}\label{gases}
\subsection{Hamiltonian and Ground State}
As an illustration of the above general result
we shall investigate the low-temperature properties
of interacting quantum gases,
i.e., the effective
dynamics of low-energy excitations above
their many-particle ground state.
We assume the microscopic dynamics of the gas to be
governed by a Hamiltonian
of the form
\begin{eqnarray}
H
&=&H_{\rm kin}+V_{2\to2}
\nonumber \\
&=&\sum_k \epsilon_k :\!a^\dagger_k a_k\!:
+{1\over4}\sum_{ijkl}
\langle lk|V|ji\rangle_\pm :\!a^\dagger_l a^\dagger_k a_j a_i\!:
\,\, ,
\label{quantum_gas}
\end{eqnarray}
with kinetic energy $H_{\rm kin}$
and a two-body interaction $V_{2\to2}$.
The single-particle energies $\epsilon_k$ include
the chemical potential.
Annihilation and creation operators obey
$[a_i,a^\dagger_j]_\mp = \delta_{ij}$
for bosons (upper sign) or fermions (lower sign),
respectively.

Each term in the Hamiltonian is normal ordered
\mbox{($:\ldots:$)} with respect to the 
noninteracting many-particle ground state.
For bosons
this ground state has all particles
in the lowest energy, zero momentum single-particle mode; 
while for fermions it
consists of a filled Fermi sea
with all momentum modes occupied up to some 
Fermi momentum $K_F$.
(For simplicity, the Fermi surface will be taken to be spherical.)
The explicit normal ordering
of the Hamiltonian is redundant
in the bosonic case.
In the fermionic case, on the other hand, it means
shuffling all operators which annihilate the fermionic vacuum
($a_i$ for states above the Fermi surface, $a^\dagger_i$ for states
below the Fermi surface) to the
right, all others
($a^\dagger_k$ for states above, $a_k$ for states below the
Fermi surface) to the left,
thereby changing sign depending on the degree of the permutation.

We will assume that, at least to a good approximation, 
the essential features of the ground state survive
even in the presence of interaction.
More specifically, we will assume that in the case of interacting
bosons the ground state still has most particles in modes
with zero, or at least very small, momentum;
and that in the case of interacting fermions 
there still exists a well-defined 
Fermi surface.
Low-energy excitations then correspond to the promotion of bosons 
from small to some slightly higher momentum,
or of fermions from just below the Fermi surface to just above it.\cite{landau}
At low temperature the regions of interest in momentum space are therefore
the vicinity of the origin (bosons) or the vicinity of the Fermi 
surface (fermions), respectively.
\subsection{Mode Elimination} 

We now wish to devise a systematic procedure for focusing onto 
these regions of interest.
To this end we consider effective theories 
(i) in the bosonic case for modes within a
sphere around the origin, of radius $\Lambda$; and
(ii) in the fermionic case for modes within a shell 
inclosing the Fermi surface, of mean radius $K_F$ and
thickness $2\Lambda$ (where $\Lambda\ll K_F$).
Whereas in the limit of large $\Lambda$ one recovers the original,
full theory, the opposite limit $\Lambda\to 0$ 
yields the desired low-energy effective theory.
In order to interpolate between these two limits
we proceed in infinitesimal steps:
we lower the 
cutoff from some given $\Lambda(s)$ to
\begin{equation}
\Lambda(s+\Delta s):=\exp(-\Delta s)\Lambda(s)
\quad,\quad \Delta s\ge 0
\label{cutoff_reduction}
\end{equation}
with $\Delta s$ infinitesimal,
thereby discarding from the theory momentum modes pertaining
to an infinitesimal shell (in the fermionic case: two shells) 
of thickness $\Delta\Lambda=\Lambda(s)\,\Delta s$;
then determine the effective dynamics of the remaining modes;
then eliminate the next infinitesimal shell, 
again determine the effective dynamics of the remaining modes,
and so on.
After each infinitesimal step we obtain a new 
effective Hamiltonian with slightly modified coupling constants.
These may also include novel couplings which had not been present
in the original theory --- in fact, the mode elimination
procedure will typically generate an infinite number of such
novel couplings.
But in many cases only a few coupling constants will change
appreciably and thus suffice to study the physical system at hand.
How these coupling constants evolve as the flow parameter
$s$ increases and hence the cutoff $\Lambda(s)$ approaches zero,
can then be described by a small set of coupled
differential equations. 
Modulo trivial scaling, these are the
{\em renormalization group equations} of the theory.

At a given cutoff $\Lambda$ the many-particle Hilbert
space (Fock space) for bosons is spanned
by the particle-free vacuum $|0_{\rm b}\rangle$ and all $n$-particle states
($n=1\ldots\infty$)
\begin{equation}
|\vec{k}_1\ldots \vec{k}_n\rangle \propto
\prod_{i=1}^n a^\dagger(\vec{k}_i) |0_{\rm b}\rangle
\quad,\quad
|\vec{k}_i|\le \Lambda
\quad,
\label{bose_excite}
\end{equation}
where the $\{{\vec{k}}_i\}$ denote the particle 
momenta and $\{a^\dagger(\vec{k}_i)\}$ the associated bosonic creation
operators.

The fermion Fock space, on the other hand, is spanned by
the filled Fermi sea (fermionic vacuum) $|0_{\rm f}\rangle$
and all its excitations which have particles above the Fermi surface
and/or missing particles (``holes'') below it,
all within a shell of thickness $2\Lambda$.
In order to cast this into a mathematical formulation
it is convenient to change coordinates,
from the true particle momenta $\{\vec{K}_i\}$ to little 
(``quasiparticle'') momenta
\begin{equation}
\vec{k}_i:=(|\vec{K}_i|-K_F)\,\hat{\vec{K}_i}
\end{equation}
and additional discrete labels
\begin{equation}
\sigma_i:=\mbox{sign}(|\vec{K}_i|-K_F)
\quad.
\end{equation}
This coordinate transformation $\vec{K}\to (\vec{k},\sigma)$
is invertible except for modes which
lie exactly on the Fermi surface.
States above the Fermi surface are labeled 
$\sigma=1$, while those below are labeled $\sigma=-1$.
The allowed excitations in fermion Fock space then
have the form ($n=1\ldots\infty$)
\begin{eqnarray}
|\vec{k}^\pm_1\ldots \vec{k}^\pm_n\rangle 
&\propto&
\prod_{i=1}^n [\theta(\sigma_i)a^\dagger(\vec{k}_i,\sigma_i)
+\theta(-\sigma_i)a(-\vec{k}_i,\sigma_i)] |0_{\rm f}\rangle
\, ,
\nonumber \\
&&\quad \quad \quad \quad \quad
|\vec{k}_i|\le \Lambda
\quad,
\label{fermi_excite}
\end{eqnarray}
where the $\{{\vec{k}}^\pm_i\}$ denote the  
momenta of particles ($+$) or holes ($-$), respectively,
and $\{a^\dagger\}$, $\{a\}$ the associated fermionic creation
and annihilation operators.
For simplicity,
we have omitted any spin quantum numbers.

It is now obvious which form the projection operator will have
that is associated with the infinitesimal cutoff reduction
(\ref{cutoff_reduction}): if applied to any of the excitations
(\ref{bose_excite}) or (\ref{fermi_excite}) 
it will simply multiply the respective state by 
a product of $\theta$ functions,
$\prod_i \theta(\Lambda-e^{\Delta s}|\vec{k}^{(\pm)}_i|)$,
to enforce the new cutoff constraint.
\subsection{Modification of the Two-Body Interaction}
Each mode elimination will yield an
effective Hamiltonian that will generally contain a slightly altered mass,
chemical potential, two-body interaction, etc., and possibly
new interactions such as an effective three-body interaction.
Here we shall restrict our attention to the
modification of the two-body interaction.
This modification is entirely due to the extra term $\Sigma$
(Eq. (\ref{sigma})) in the effective Hamiltonian,
\begin{eqnarray}
\Sigma
&=&
-{1\over32\hbar}\left\{
\sum_{abcd} \sum_{ijkl}
\langle lk|V|ji\rangle_\pm \langle dc|V|ba\rangle_\pm 
\right.
\nonumber \\
&& 
\left. \times
P:\!a^\dagger_l a^\dagger_k a_j a_i\!: Q{1\over {\cal L}^{(0)}} 
Q:\!a^\dagger_d a^\dagger_c a_b a_a\!: P
+\mbox{h.c.}
\right\}
\,\, ,
\label{full_sigma}
\end{eqnarray}
where to the given order in perturbation theory 
${\cal L}^{(0)}$ just coincides with ${\cal L}_{\rm kin}$.
The two projectors $P$ at both ends of the operator product
ensure that all external momenta lie below
the new, reduced cutoff;
whereas the projectors $Q$ in the center force at least one internal momentum
to lie in that infinitesimal shell which has just been eliminated.
Therefore, at least one pair of field operators must pertain
to the eliminated $Q$-modes, and hence be contracted.
The product of the remaining six field operators can then be rearranged
with the help of Wick's theorem to yield a
decomposition
\begin{equation}
\Sigma=\Sigma_6+\Sigma_4
+\Sigma_2+\Sigma_0
\quad,
\end{equation}
each $\Sigma_n$ being a normal ordered product of
$n$ field operators.
The various terms shift the ground state energy ($n=0$),
modify the mass, the chemical potential, or more generally the form of the
single-particle dispersion relation ($n=2$), modify the
two-body interaction ($n=4$), and generate a new effective
three-body interaction ($n=6$).

As we want to focus on the modification of the two-body interaction,
we consider only the term with $n=4$.
Neglecting the energy of the external modes
we find for bosons ($\hbar=1$)
\begin{equation}
\Delta \langle lk|V|ji\rangle_+
= -\Delta \left[\sum_{ab} {1\over 2(\epsilon_a+\epsilon_b)}
\langle lk|V|ba\rangle_+ \langle ab|V|ji\rangle_+\right]
\label{mod_bosons}
\end{equation}
and for fermions
\begin{equation}
\Delta \langle lk|V|ji\rangle_-
= \Delta^{(\rm ZS)}_{lk|ji} + \Delta^{(\rm ZS')}_{lk|ji} 
+ \Delta^{(\rm BCS)}_{lk|ji}
\label{fermion_mod}
\end{equation}
with
\begin{eqnarray}
\Delta^{(\rm ZS)}_{lk|ji}
&=& -\Delta [\sum_{ab}
{\theta(\sigma_a)\theta(-\sigma_b)-\theta(-\sigma_a)\theta(\sigma_b)
\over \epsilon_a-\epsilon_b}
\nonumber \\
&&
\quad
\times
\langle la|V|bi\rangle_- \langle bk|V|ja\rangle_-
]
\quad,
\label{zs}
\end{eqnarray}
its cross term
\begin{equation}
\Delta^{(\rm ZS')}_{lk|ji} = - \Delta^{(\rm ZS)}_{kl|ji}
\quad,
\label{zs_prime}
\end{equation}
and
\begin{eqnarray}
\Delta^{(\rm BCS)}_{lk|ji}
&=& 
-\Delta [\sum_{ab}
{\theta(\sigma_a)\theta(\sigma_b)
- \theta(-\sigma_a)
\theta(-\sigma_b)
\over 2(\epsilon_a+\epsilon_b)}
\nonumber \\
&&
\quad
\times
\langle lk|V|ba\rangle_- \langle ab|V|ji\rangle_-
]
\quad.
\label{bcs}
\end{eqnarray}
The $\Delta$ in front of the sums signifies that
at least one of the internal modes ($a,b$) must 
lie in the eliminated shell.
In the bosonic case the modification of the two-body
interaction can be associated with a 1-loop ``ladder'' diagram.
In the fermionic case, on the other hand, 
there are three distinct contributions
which, with hindsight, may be identified with 
``zero sound'' (ZS,ZS') and BCS diagrams.\cite{shankar}
The ZS contribution and its cross term ZS' account for particle-hole
excitations ($\sigma_a=\pm 1,\sigma_b=\mp 1$),
while the BCS term describes 2-particle
($\sigma_a=\sigma_b=+1$) or 2-hole ($\sigma_a=\sigma_b=-1$)
excitations.

The above formulae serve as the starting point for the investigation
of a variety of specific physical systems.
With their help one can derive 
such diverse results as the 1-loop renormalization group equation
for an interacting
Bose gas, the 1-loop $\beta$ function of $\phi^4$ theory, the
screening of fermion-fermion interactions, or 
the BCS instability.
Details of these applications are presented 
in the following section.
\section{Examples}\label{examples}
\subsection{Bosons with point interaction}
For spinless bosons with point interaction
($\delta$ function potential in real space) it is
\begin{equation}
\langle lk|V|ji\rangle_+
={2U\over\Omega}\, \delta_{\vec{k}_i+\vec{k}_j,\vec{k}_k+\vec{k}_l}
\quad,
\label{point_coupling}
\end{equation}
with the Kronecker symbol enforcing momentum conservation,
$\Omega$ being the spatial volume,
and $U$ the coupling constant.
Provided the magnitude of the external momenta $\vec{k}_i$, $\vec{k}_j$ 
is negligible compared to
the cutoff $\Lambda$, 
momentum conservation implies that
the internal modes $a,b$ must {\em both} lie in the
eliminated shell, and that hence 
$\epsilon_a=\epsilon_b=\epsilon_\Lambda$.
Application of the general formula (\ref{mod_bosons}) then yields
\begin{equation}
\Delta U= -{U^2 \over 2\Omega\epsilon_\Lambda}
\Delta \left[
\sum_{|\vec{k}_a|,|\vec{k}_b|\le\Lambda}
\delta_{\vec{k}_i+\vec{k}_j,\vec{k}_a+\vec{k}_b}
\right]
\quad,
\label{point_mod}
\end{equation}
where the sum
\begin{eqnarray}
\Delta \left[
\sum_{|\vec{k}_a|,|\vec{k}_b|\le\Lambda}
\delta_{\vec{k}_i+\vec{k}_j,\vec{k}_a+\vec{k}_b}
\right]
&\approx& 
\Delta\left[\sum_{|\vec{k}_a|,|\vec{k}_b|\le\Lambda}
\delta_{\vec{k}_b,-\vec{k}_a}\right]
\nonumber \\
&=& 
\sum_{|\vec{k}_a|\in [\Lambda-\Delta\Lambda,\Lambda]} 1
\end{eqnarray}
simply counts the number of eliminated states.
For a spherical cut in momentum space this number
of states is given by
\begin{equation}
\sum_{|\vec{k}_a|\in [\Lambda-\Delta\Lambda,\Lambda]} 1 =
\rho(\epsilon_\Lambda)\,{d\epsilon_\Lambda\over d\Lambda}\,
\Delta\Lambda
\quad,
\label{number_states}
\end{equation}
with $\rho(\epsilon_\Lambda)$ denoting the density of states at
the cutoff.
With $\Delta\Lambda=\Lambda\,\Delta s$ we thus obtain the
flow equation
\begin{equation}
\Delta U= 
-{d\ln\epsilon_\Lambda \over d\ln\Lambda}\,
{\rho(\epsilon_\Lambda)\over 2\Omega}\,
U^2\cdot \Delta s
\quad.
\label{boson_flow}
\end{equation}
 
For a dilute gas of nonrelativistic bosons 
in three spatial dimensions, with mass $m$,
dispersion relation $\epsilon_\Lambda=\Lambda^2/2m$
and density of states $\rho(\epsilon_\Lambda)=\Omega m\Lambda/2\pi^2$
the flow equation reduces to
\begin{equation}
\Delta U= - {m\Lambda \over 2\pi^2} U^2\cdot \Delta s
\quad.
\label{u_flow}
\end{equation}
By its very definition the sequence of effective theories retains
complete information about the system's low-energy dynamics.
Observables pertaining to this low-energy
dynamics 
are therefore unaffected by the successive mode elimination, and
hence independent of $s$.
For example, the scattering length\cite{abrikosov}
\begin{equation}
{a} = {m\over 4\pi}\left[ U(s) - U(s)^2 \int_{|\vec{p}|\le\Lambda(s)} 
{d^3p\over(2\pi)^3}\, {m\over \vec{p}^2} \right]
\end{equation}
stays constant under the flow (\ref{u_flow}), as
the $s$-dependence of the parameters $U$ and $\Lambda$
just cancels out
(up to third order corrections).
\subsection{The link to $\phi^4$ theory} 
There is an interesting relationship between the result 
(\ref{boson_flow})
and the 1-loop $\beta$ function for real $\phi^4$ theory.
The $\phi^4$ Hamiltonian describes the dynamics of coupled anharmonic
oscillators. It reads, in three spatial dimensions,
$H=H^{(0)} + V$
with kinetic energy
\begin{eqnarray}
H^{(0)}
&=&
{1\over2}\int d^{3}x :\!\left[\pi(x)^2
+ |\nabla\phi(x)|^2 + m^2\phi(x)^2\right]\!:
\nonumber \\
&=&
\sum_{\vec k} \epsilon_{\vec k}\, a^\dagger_{\vec k} a_{\vec k}
\end{eqnarray}
and interaction
\begin{eqnarray}
V
&=&
{g\over 4!} \int d^{3}x\,\phi(x)^4
\nonumber \\
&=&
{g\over 4!\Omega}
\sum_{\vec{k}_1\vec{k}_2\vec{k}_3\vec{k}_4}
\prod_{\alpha=1}^4
{1\over\sqrt{2\epsilon_{\vec{k}_\alpha}}}
(a_{\vec{k}_\alpha}+a^\dagger_{-\vec{k}_\alpha})
\delta_{\sum \vec{k}_i,0}
\, .
\end{eqnarray}
Here $m$ denotes the mass, 
$\Omega$ the spatial volume, $g$ the coupling constant,
and $\epsilon_{\vec k}$ the single-particle energy
\begin{equation}
\epsilon_{\vec{k}}=\sqrt{\vec{k}^2+m^2}
\quad.
\end{equation}
The field $\phi$ and its conjugate momentum $\pi$ 
are time-independent (Schr\"odinger picture) operators which satisfy the
commutation relations for bosons, and $a$, $a^\dagger$ are the
associated annihilation and creation operators.
While the kinetic part of the Hamiltonian is normal ordered
($:\ldots :$), the interaction is not.
 
When expressed in terms of annihilation and creation
operators 
the Hamiltonian takes on a form which is very similar to that of the
quantum gas Hamiltonian (\ref{quantum_gas}).
More precisely, the $\phi^4$ Hamiltonian {\em contains}
a Bose gas Hamiltonian with two-body interaction matrix element
\begin{equation}
\langle lk|V|ji\rangle_+
=\left({4\atop 2}\right)
{g\over 4!\Omega\sqrt{\epsilon_i\epsilon_j\epsilon_k\epsilon_l}}
\delta_{\vec{k}_i+\vec{k}_j,\vec{k}_k+\vec{k}_l}
\quad.
\end{equation}
The derivation of a flow
equation for $g$ can now proceed in the same vein as that for $U$,
again starting from Eq. (\ref{mod_bosons}).
Now, however, apart from $2\to2$ particle scattering the
$\phi^4$ Hamiltonian with its novel interactions
$a^\dagger a^\dagger a^\dagger a$, $a^\dagger a^\dagger a^\dagger a^\dagger$
etc.
also permits $2\to4$ and $2\to6$ scattering.
Therefore, in Eq. (\ref{mod_bosons}) the intermediate state
may be not just $|ab\rangle$, but also
$|abik\rangle$, $|abil\rangle$, $|abjk\rangle$, $|abjl\rangle$
or $|abijkl\rangle$.
As long as the magnitude of the external momenta is negligible
compared to the cutoff, it is in all six cases $\epsilon_a=
\epsilon_b=\epsilon_\Lambda$ and 
\begin{equation}
\langle lk|V|\ldots\rangle_+ \langle\ldots |V|ji\rangle_+ =
{g\over 4\Omega\epsilon_\Lambda^2}
\langle lk|V|ji\rangle_+ \delta_{\vec{k}_b,-\vec{k}_a}
\,\, .
\end{equation}
Hence in order to account for the larger set of allowed intermediate states
we merely have to
introduce an extra factor $6$, and obtain thus
\begin{equation}
\Delta g= 
-{d\ln\epsilon_\Lambda \over d\ln\Lambda}\,
{3\rho(\epsilon_\Lambda)\over 8\Omega\epsilon_\Lambda^2}\,
g^2\cdot \Delta s
\quad.
\end{equation}
For $\Lambda\gg m$ it is
$\epsilon_\Lambda=\Lambda$, 
$\rho(\epsilon_\Lambda)=\Omega\epsilon_\Lambda^2/2\pi^2$, and the
flow equation reduces to
\begin{equation}
\Delta g= -{3g^2\over 16\pi^2}\Delta s
\quad,
\end{equation}
in agreement with the well-known 1-loop result for the
$\beta$ function of $\phi^4$ theory.\cite{fisher,generalize}
\subsection{Screening of fermion-fermion interactions}
We consider nonrelativistic fermions in spatial
dimension $d$ ($d\ge 2$) which interact through a two-body interaction
\begin{eqnarray}
\langle lk|V|ji \rangle_- 
&=& 
[V(\vec{q})\delta_{s_l s_i}\delta_{s_k s_j}
- V(\vec{q}')\delta_{s_k s_i}\delta_{s_l s_j}]
\nonumber \\
&&
\,\times
\delta_{\vec{K}_i+\vec{K}_j,\vec{K}_k+\vec{K}_l}
\quad,
\label{screen}
\end{eqnarray}
duly antisymmetrized to account for Fermi statistics,
and with $\{s_\alpha\}$ denoting the spin quantum numbers
and $\vec{q}, \vec{q}'$ 
the respective momentum transfers
\begin{eqnarray}
\vec{q}&:=&\vec{K}_l-\vec{K}_i=\vec{K}_j-\vec{K}_k
\quad,
\nonumber \\
\vec{q}'&:=&\vec{K}_k-\vec{K}_i=\vec{K}_j-\vec{K}_l
\quad.
\end{eqnarray}
We investigate scattering processes for which
\begin{equation}
0 < |\vec{q}|,\Lambda \ll |\vec{q}'|, |\vec{K}_i+\vec{K}_j|, K_F
\quad.
\end{equation}
In this regime only the ZS contribution (\ref{zs}) can
significantly modify the two-body interaction;
its cross term ZS' (Eq. (\ref{zs_prime})),
as well as the BCS contribution (\ref{bcs}),
are suppressed by a factor $\Lambda/K_F$.
This can be seen directly from the geometry of the 
Fermi surface.
The three constraints on the intermediate state:
(i) both $\vec{K}_a$ and $\vec{K}_b$ 
lie in the cutoff shell of thickness $2\Lambda$;
(ii) more stringently, one of them lie in the infinitesimal shell to
be eliminated; and 
(iii) $\vec{K}_a-\vec{K}_b=-\vec{q}'$ (ZS') or
$\vec{K}_a+\vec{K}_b=\vec{K}_i+\vec{K}_j$ (BCS), respectively ---
reduce the momentum space volume available to the internal momentum
$\vec{K}_a$ to $O(K_F^{d-2}\Lambda\Delta\Lambda)$.
In contrast, for $|\vec{q}|\sim \Lambda$ 
the ZS contribution
with its condition $\vec{K}_a-\vec{K}_b=-\vec{q}$ 
allows a momentum space volume
of the order $K_F^{d-1}\Delta\Lambda$.

To evaluate the ZS contribution 
at some given momentum transfer $\vec{q}$,
we first  define the angle $\vartheta$ between $-\vec{q}$ and
the internal momentum $\vec{K}_a$,
\begin{equation}
\cos\vartheta\equiv z:=-{\vec{q}\cdot \vec{K}_a \over |\vec{q}||\vec{K}_a|}
\quad,
\end{equation}
change coordinates from original ($\vec{K}$) to little ($\vec{k}$) momenta,
and write, up to corrections of order $|\vec{q}|/K_F$,
\begin{eqnarray}
\epsilon_a-\epsilon_b = v_F (|\vec{K}_a|-|\vec{K}_b|)
&=& v_F (\sigma_a |\vec{k}_a| -\sigma_b |\vec{k}_b|)
\nonumber \\
&=& v_F |\vec{q}| z
\end{eqnarray}
with $v_F$ denoting the Fermi velocity.
Next we note that the term with $\theta(\sigma_a)\theta(-\sigma_b)$
and the term with $\theta(-\sigma_a)\theta(\sigma_b)$ yield identical
contributions; therefore it suffices to consider only
the first term and then multiply it by two.
Finally, assuming that in the interaction matrix element (\ref{screen})
the cross term is negligible,
\begin{equation}
|V(\vec{q}')| \ll
|V(\vec{q})|
\quad,
\end{equation}
the two matrix elements in Eq. (\ref{zs}) can simply be replaced
by $V(\vec{q})^2$ modulo Kronecker symbols for spin and momentum
conservation.
By virtue of these Kronecker symbols one
of the two summations over internal modes collapses trivially,
leaving
\begin{eqnarray}
\Delta V(\vec{q})
&=& 
-{2\over v_F} \Delta\left[
\sum_a
{\theta(\Lambda-|\vec{k}_a|)\theta(|\vec{k}_a|-|\vec{q}|z+\Lambda)}
\right.
\nonumber \\
&&
\left.
\quad\quad\quad
\times
{{\theta(\sigma_a)\theta(|\vec{q}|z -|\vec{k}_a|)}
\over |\vec{q}|z}
\right]
\cdot V(\vec{q})^2
\quad.
\end{eqnarray}
Here the first two $\theta$ functions explicitly enforce
the sharp cutoff constraint for both $\vec{k}_a$ and $\vec{k}_b$
($|\vec{k}_a|,|\vec{k}_b|\le\Lambda$),
while the latter two
$\theta$ functions enforce $\sigma_a=1$
and $\sigma_b=-1$, respectively.
Under these constraints it is always $\vartheta\in [0,\pi/2)$
and hence $z>0$.

The above equation can be immediately integrated from cutoff
$\Lambda\gg |\vec{q}|,|\vec{k}_a|$ (symbolically,
$\Lambda\to\infty$)
down to
$\Lambda\ll |\vec{q}|,|\vec{k}_a|$
(symbolically, $\Lambda\to 0$),
to yield the total modification of the two-body interaction:
\begin{eqnarray}
\lefteqn{{1\over V_{\rm eff}(\vec{q})}
-{1\over V_{\rm bare}(\vec{q})}
=}
\nonumber \\
&&
{2\over v_F}\!
\left.
\sum_a
{\theta(\Lambda-|\vec{k}_a|)\theta(|\vec{k}_a|-|\vec{q}|z+\Lambda)
\theta(\sigma_a)\theta(|\vec{q}|z -|\vec{k}_a|)
\over |\vec{q}|z}
\right|_{0}^{\infty}
\nonumber \\
&& \mbox{}
\end{eqnarray}
At the lower bound ($\Lambda\to 0$) the various conditions imposed by the
$\theta$ functions cannot all be satisfied simultaneously, and therefore
the product of $\theta$ functions vanishes.
At the upper bound ($\Lambda\to\infty$), 
on the other hand, the cutoff constraints
imposed by the first two $\theta$ functions are trivially
satisfied and thus can be omitted.
In this case the sum over $a$ is evaluated by turning it 
into two integrals, one over a radial
variable such as $|\vec{k}_a|$ or $\epsilon_a$, the other
over the solid angle.
At a given solid angle and hence given $z$, the fourth
$\theta$ function restricts the radial
integration to the range $|\vec{k}_a|\in [0,|\vec{q}|z]$ or,
equivalently,
$\epsilon_a\in [0,v_F |\vec{q}|z]$.
This energy interval in turn corresponds to a number
$[\rho(\epsilon_F) v_F |\vec{q}| z]$ of states,
$\rho(\epsilon_F)$ being the density of states
at the Fermi surface.
(It includes a factor to account for the spin degeneracy.)
The integration over the solid angle
is constrained to a semi-sphere, due to $\vartheta\in[0,\pi/2)$,
and hence reduced by a factor $1/2$ as compared to a full-sphere
integration.
Altogether we obtain
\begin{equation}
{1\over V_{\rm eff}(\vec{q})}
-{1\over V_{\rm bare}(\vec{q})}
=
{2\over v_F}
\,\, {1\over2}\,\,
\rho(\epsilon_F) v_F |\vec{q}|z\,
{1\over |\vec{q}|z}
=
\rho(\epsilon_F)
\end{equation}
and thus
\begin{equation}
{V_{\rm eff}(\vec{q})}=
\left[
{1\over V_{\rm bare}(\vec{q})}+ \rho(\epsilon_F)
\right]^{-1}
\quad.
\end{equation}
This result describes the well-known screening of
fermion-fermion interactions.\cite{fetter}
\subsection{BCS instability}
Our last example pertains to fermions with an attractive pairing
interaction
\begin{equation}
\langle lk|V|ji \rangle_- = -V\, \delta_{\vec{K}_j,-\vec{K}_i}
\delta_{\vec{K}_l,-\vec{K}_k}
[\delta_{s_l s_i}\delta_{s_k s_j} - \delta_{s_k s_i}\delta_{s_l s_j}]
\, ,
\end{equation}
which is the simplest form of BCS theory.\cite{schrieffer}
Due to the pairing condition $\vec{K}_j=-\vec{K}_i$,
$\vec{K}_l=-\vec{K}_k$
it is impossible to satisfy
in the ZS and ZS' terms the requirement that at least one
of the internal modes be in the eliminated shell.
Hence only the BCS term (\ref{bcs}) can modify
the coupling constant.
In the BCS term there are contributions with $\theta(\sigma_a)\theta(\sigma_b)$
and $\theta(-\sigma_a)\theta(-\sigma_b)$, respectively, which yield identical
results; therefore it suffices to consider only
the first contribution and then multiply it by two.
The pairing condition implies $\epsilon_a=\epsilon_b=\epsilon_\Lambda$,
which for modes in the upper ($\sigma=+1$)
eliminated shell is given by $\epsilon_\Lambda=v_F \Lambda$.
The eliminated shell itself
covers an infinitesimal energy interval of width
$[v_F \Lambda\Delta s]$,
which in turn corresponds to a number
$[\rho(\epsilon_F) v_F \Lambda\Delta s]$ of states.
Of the two summations over internal modes one 
collapses trivially due to momentum and spin conservation,
leaving
\begin{eqnarray}
\Delta V
&=&
{V^2\over 2v_F\Lambda}
\Delta \left[\sum_{a} {\theta(\sigma_a)}\right]
\nonumber \\
&=&
{V^2\over 2v_F\Lambda}\,
\rho(\epsilon_F)\, v_F \Lambda\Delta s
= {\rho(\epsilon_F)\over 2}\,V^2 \cdot \Delta s
\quad.
\end{eqnarray}
 
From this flow equation for the BCS coupling $V(s)$
we immediately conclude that 
as long as the initial coupling $V(0)$ is positive,
$V(s)$ diverges as $s\to\infty$.
This indicates the occurence of binding (``Cooper pairs'')
at very low temperatures.
Furthermore, we can again convince ourselves that
the sequence of effective theories retains complete information
about the system's low-energy dynamics:
low-energy observables
such as the zero-temperature gap\cite{schrieffer}
\begin{equation}
\Delta_0 =
2\Lambda(s) \exp\left[
-{2\over \rho(\epsilon_F) V(s)}\right]
\end{equation}
do not depend on the flow parameter $s$
and are thus unaffected by the successive mode elimination.
\section{Conclusion}\label{summary}
We have succeeded in linking renormalization
to transport theory.
Our line of argument proceeded from the exact microscopic dynamics,
via the Nakajima-Zwanzig projection technique, to a
macroscopic transport equation for selected expectation values,
and then,
via second order perturbation theory, Markovian limit,
choice of the Robertson projector and a suitable
rearrangement of terms in the transport equation,
to the approximate effective Hamiltonian which governs
the non-dissipative part of the macroscopic dynamics.
We investigated the ramifications of this result for the
low-energy dynamics of interacting quantum gases:
contracting the set of selected expectation values by 
discarding iteratively 
those observables which pertain to short-wavelength
excitations,
we obtained a sequence of effective Hamiltonians which describe
the dynamics on successively larger length scales.
We then focused on
the two-body interaction in these effective
Hamiltonians 
and convinced ourselves 
--in several rather diverse applications-- that
it varies in accordance with 1-loop
renormalization group equations.
We have thus 
substantiated our original claim 
that renormalization group equations can be obtained
within the
Nakajima-Zwanzig projection approach,
and that hence renormalization can be embedded into the
general mathematical framework of transport theory.

There remain many open questions worth investigating.
Clearly, a unified theoretical framework for dissipative transport
and renormalization will have to prove its merits in new applications
where the conventional approaches fail:
for example, when time scales are no longer well separated
and the Markovian limit ceases to be justified;
or when in the course of successive mode elimination
one starts to discard states with a finite population,
thus introducing
dissipation into the effective macroscopic dynamics.
It is my hope that the present paper will help stimulate
research efforts in these and related directions.
%
%


\begin{references}
\bibitem{generalize}
A more detailed derivation of the renormalization group equations
for $\phi^4$ theory by means of projection techniques,
as well as the generalization to arbitrary
spatial dimension, can be found in Ref. \CITE{muller:rau}.
\end{references}
\end{document}